\documentclass[12pt,dvips,a4paper]{article}
\usepackage{epsfig}
\usepackage{chicnarm}

\newcommand{\JEM}{Journal of Engineering Mechanics, ASCE}

\newcommand{\IJNME}{International Journal for Numerical Methods in Engineering}

\newcommand{\EFM}{Engineering Fracture Mechanics}
\newcommand{\IJF}{International Journal of Fracture}

\newcommand{\MS}{Materials and Structures}

\newcommand{\AES}{Advances in Engineering Software}

\newcommand{\sbf}[1]{\boldsymbol{#1}}

\newcommand{\veps}{\varepsilon}

\newcommand{\ignore}[1]{}

\newcommand{\optional}[1]{}

\usepackage{amssymb}
\usepackage{amsmath}
 \usepackage{setspace}

\begin{document}

\renewcommand{\title}{Meso-scale approach to modeling concrete subjected to thermo-mechanical loading}
\newcommand{\authors}{Peter Grassl\footnote{Corresponding author. Email: grassl@civil.gla.ac.uk, Department of Civil Engineering, University of Glasgow, G12 8LT Glasgow, United Kingdom.} and Chris Pearce\footnote{Department of Civil Engineering, University of Glasgow, G12 8LT Glasgow, United Kingdom.}}

\begin{center} \begin{LARGE} \textbf{\title} \end{LARGE} \end{center}

\begin{center}
\authors\\

\end{center}

\begin{center}
Accepted in Journal of Engineering Mechanics-ASCE\\
30th October 2009
\end{center}

CE Database subject headings: Concrete, Cracking, Thermal analysis, Plasticity

\section*{Abstract}
Concrete subjected to combined compressive stresses and temperature
loading exhibits compressive strains, which are considerably greater
than for concrete subjected to compressive stresses alone. This
phenomenon is called transient thermal creep or load induced thermal
strain and is usually modeled by macroscopic phenomenological
constitutive laws which have only limited predictive capabilities.
In the present study a meso-scale modeling approach is proposed in
which the macroscopically observed transient thermal creep results
from the mismatch of thermal expansions of the meso-scale
constituents. The meso-structure of concrete is idealized as a
two-dimensional three phase material consisting of aggregates,
matrix and interfacial transition zones (ITZ). The nonlinear
material response of the phases are described by a damage-plasticity
interface model. The meso-scale approach was applied to analyze
compressed concrete specimens subjected to uniform temperature
histories and the analysis results were compared to experimental
results reported in the literature. 

\section*{Introduction}

The mechanical behaviour of concrete is significantly altered
when exposed to high temperatures, whereby properties such as
strength and stiffness are generally found to decrease with
temperature and time dependent creep characteristics are found to be
significantly enhanced. Experimental evidence also suggests that
strength, stiffness and creep depend on the combined mechanical and
thermal loading history.

When an unloaded concrete specimen is subjected to transient high
temperatures it undergoes thermal expansion. When the same test is
performed under sustained mechanical load, the total strains, minus
the elastic strains, are significantly different from those observed
in the unloaded case \cite{KhoGraSul85,The87,Sch88a}. This
difference in the strains is normally referred to as transient
thermal creep or load induced thermal strain. It is worth noting
that the use of the word 'creep' is a subject for debate, since
transient thermal creep is not time-dependent, as is the case for
other forms of creep; for consistency with other work on this
subject, the term 'creep' will continue to be used in this paper.

Transient thermal creep is an irreversible thermo-mechanical strain,
only observed on first heating and is considered a
quasi-instantaneous response \cite{KhoGraSul85,The87}.
It is generally recognized that its presence provides an important
relaxation of compressive stresses when concrete is heated beyond
$100$~$^\circ$C.

However, the irreversibility of transient thermal creep can lead to
the build-up of potentially critical tensile strains in restrained
concrete members during cooling \cite{NiePeaBic02}. Recent
experimental studies by Colina et al.
\cite{ColSer04,HasCol06a,SabMefColPla08} performed at very slow
heating rates up to temperatures of $400$~$^\circ$C have provided
further confirmation of the existence of transient thermal creep as
well as the principal characteristics outlined above. They also
hypothesized that the phenomenon depends only on irrecoverable
physical and chemical processes that take place due to increase of
temperature. \cite{ColSer04} also identified that transient thermal
creep is less in ordinary concrete (OC), which is characterized by
higher permeability, compared to that of high performance concrete
(HPC) with lower permeability. \cite{HasCol06a,SabMefColPla08}
concluded that dehydration of the cement paste is one of the
critical processes in the development of transient thermal creep.

There have been a number of efforts to phenomenologically model
transient thermal creep strains during combined thermal and
mechanical loading. Thelandersson \cite{The82} and Schneider
\cite{Sch88a} suggested uniaxial expressions for the transient
thermal creep, with the former applying the theory of plasticity and
extending the model to 3D. Khennane and Baker
\cite{KheBak92,KheBak93} also adopted the theory of plasticity and
based their numerical implementation of transient thermal creep on
the work of de Borst and Peeters \cite{BorPee89}. Thelandersson
\cite{The87} proposed a general 3D model that has been adopted by a
number of authors (e.g. \cite{NecMefRey02}), but it is frequently
unclear how the critical transient thermal creep parameter has been
determined. Nielsen et al. \cite{NiePeaBic02} derived the transient
thermal creep parameter from experimental data reported by Schneider
\cite{Sch88a} and Khoury et al. \cite{KhoGraSul85}. This model was
subsequently implemented into a continuum Finite Element formulation
by Pearce et al. \cite{PeaNieBic04}. Gawin et al.
\cite{GawPesSch04,GawPesSch06} adopted a similar approach, although
they related the transient thermal creep strains to a
thermo-chemical damage parameter rather than temperature. Ba\v{z}ant
and Chern \cite{BazChe87} extended their original creep model to
include the effect of temperature alterations, suggesting that both
heating and cooling result in transient thermal creep. Thienel and
Rostasy \cite{ThiRos96} proposed a general constitutive model for
transient thermal strains based on their own experimental biaxial
results. Youssef and Moftah \cite{YouMof07} have undertaken a
comparison of a number of phenomenological models for transient
thermal creep.

In general, the mechanics of concrete subjected to temperature
loading are not well understood and the macroscopic phenomenological
constitutive models which are based on existing experimental results
cannot be relied upon to predict the structural response outside the
range of these experiments. Furthermore, there is still dispute
about the importance of moisture, drying shrinkage and differences
in the thermal properties of the constituents of the meso-structure.
The authors believe that a detailed study of the meso- and
micro-structural behavior of concrete can help in resolving this
matter by allowing the mechanisms which dominate the macroscopic
response to be isolated. Recent promising research in this direction
explored the macroscopic transient thermal creep of concrete via a
two-phase meso-scale modeling approach which considered the thermal
expansion and the thermal damage of the elastic properties
\cite{WilRheShi04,WilRheXi05,GroDumHamMou07}. 

In the present work, a different meso-scale description is adopted,
in which the meso-structure is enhanced by an additional phase,
which represents the interfacial transition zone (ITZ) between the
aggregates and the matrix. The macroscopic transient thermal creep
is described by the nonlinear mechanical response, which is caused
by the mismatch of thermal expansion of the different phases. The
aim of the present study is to show that this mismatch of expansion has an important influence on the macroscopically observed transient thermal creep. 

The long term aim of the present research is to improve the
understanding of the macroscopic response of concrete subjected to
temperature loading by investigating the influence of different
processes on the meso and micro-scale. The present manuscript is
focused on the nonlinear response due to the mismatch of thermal
expansion on the meso-scale. Further work will be required to
investigate other influences, e.g. moisture, creep and relaxation on
micro and meso-scale. The present study is only a first step towards
an understanding of the mechanisms that cause transient thermal
creep.

Meso-scale modeling of strongly heterogeneous materials like
concrete is computationally very demanding. Continuum formulations
for the description of fracture, especially at the interface between
materials of significantly different stiffness, is difficult and can
be accompanied by severe numerical problems. Therefore, a lattice
approach is applied in the present study
\cite{BolSai98,Kaw77,MorSawKob93}. The heterogeneity of the
material is described by spatially varying material properties of
the lattice elements with respect to their position within the
meso-structure \cite{SchMie92b,LilMie03}, which is composed of
aggregates, matrix and ITZs.
It is recognised that each of the three phases comprise a complex microstructure.
However, in the present study the response of these phases is described by phenomenological constitutive models.
Alternatively, for at least some of these phases, micromechanics approaches could be used \cite{BudOco76,HeuLemUlm05,DorKonUlm06}.

\section*{Modeling approach}
The present approach to modeling the response of concrete subjected
to combined thermal and mechanical loading is based on a meso-scale
description. Aggregates, matrix and interfacial transition zones are
modeled as separate phases with different material properties. A
lattice approach is used to discretize the domain. For each lattice
element, a nonlinear stress-strain relationship based on a
combination of damage and plasticity is used.
This damage-plasticity constitutive model was originally proposed for the modeling of concrete subjected to cyclic loading \cite{GraRem08} and is extended in the present study to describe the influence of thermal expansion.

\subsection*{Lattice framework}\label{sec:lattice}
The domain is decomposed into polygons based on the Voronoi tessellation \cite{Aur91}.
The nodes for the Voronoi tessellation are placed sequentially in the domain, whereby the coordinates of each node are determined randomly.
A minimum distance $d_{\rm m}$ is enforced iteratively between the nodes.
For each randomly placed node the distance to the existing nodes is checked. Only if the smallest distance between two nodes is greater than the prescribed minimum distance, is the node accepted \cite{ZubBaz87}.

\begin{figure}[h!]
\begin{center}
\begin{tabular}{cc}
\epsfig{file = 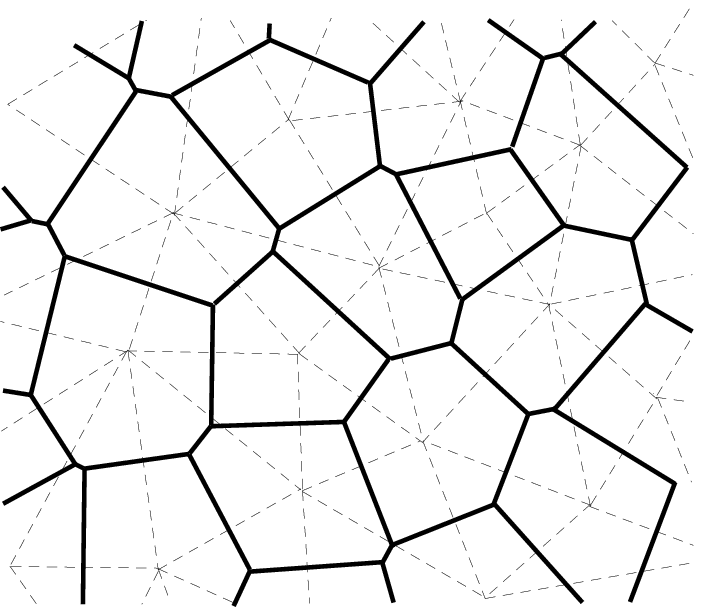, width =5.cm} & \epsfig{file = 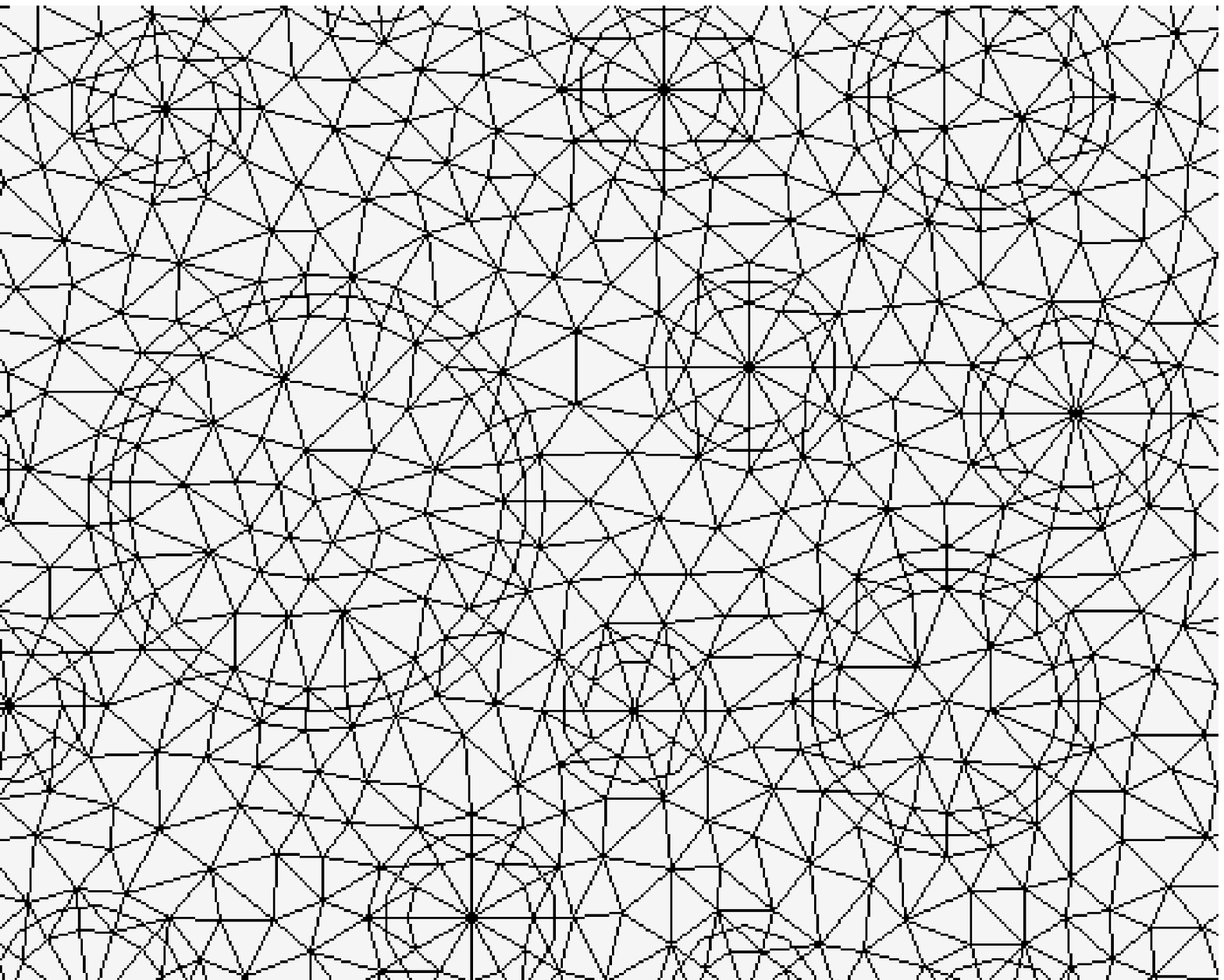, width =5.5cm}\\
(a) & (b)
\end{tabular}
\end{center}
\caption{(a) Voronoi tessellation and dual Delaunay triangulation. Solid lines represent the cells defined by the Voronoi tessellation. Dashed lines represent the lattice elements describing the interaction between the nodes. (b) Meso-structure of concrete modeled by placing lattice elements perpendicular to the interfaces between aggregates and matrix.}
\label{fig:vorDel}
\end{figure}

For this iterative process, the relationship between the number of nodes for a chosen domain and the minimum distance determines the distribution of the distances between nodes.
The number of nodes for a minimum distance and domain size $A_{\rm d}$ is expressed in the form of a density
\begin{equation}
\rho = \dfrac{n d_{\rm m}^2}{A_{\rm d}}
\end{equation}
If the density of the nodes is small, the variation of Voronoi cell sizes is large. On the other hand, if the domain is
saturated with nodes, i.e. maximum number of nodes for a specific
domain and minimum distance, the Voronoi cells are of similar size. Bolander and Saito
\cite{BolSai98} determined numerically the density for a saturated
node arrangement to be $\rho \approx 0.68$. The dual to the Voronoi
tessellation is the Delaunay triangulation shown in
Figure~\ref{fig:vorDel}a. The edges of the Delaunay triangulation
are the lattice elements, which describe the interaction between the
nodes.

The use of an irregular lattice for the description of the fracture process of quasi-brittle materials is important, since the fracture patterns are strongly influenced by the alignment of the lattice \cite{JirBaz95,SchGar96}.
Related problems are known from continuum fracture approaches, where mesh alignment can have a strong influence on the fracture patterns obtained \cite{GraJir04c,GraRem07,JirGra08}.

The meso-structure of concrete is discretized by placing lattice
elements perpendicular to the interface between aggregates and the
matrix. In Figure~\ref{fig:vorDel}b, a schematic meso-structure and
the corresponding discretization is shown.

The distribution of the aggregates is determined randomly using a
cumulative distribution function \cite{CarCorPuz04}. A standard
pseudo-random number generator is used to generate probabilities
from which the diameter $d$ is determined. This procedure is
repeated until the chosen volume fraction of aggregates is obtained.
In the present 2D idealisation, aggregates are idealized as
cylindrical inclusions of diameter $d$. The aggregates are placed
sequentially by means of randomly generated coordinates within the
area of the specimen. For each set of generated coordinates it is
checked that no overlap with existing aggregates occurs. However,
overlaps with boundaries are permitted.

\begin{figure}[h!]
\begin{center}
\begin{tabular}{cc}
\epsfig{file=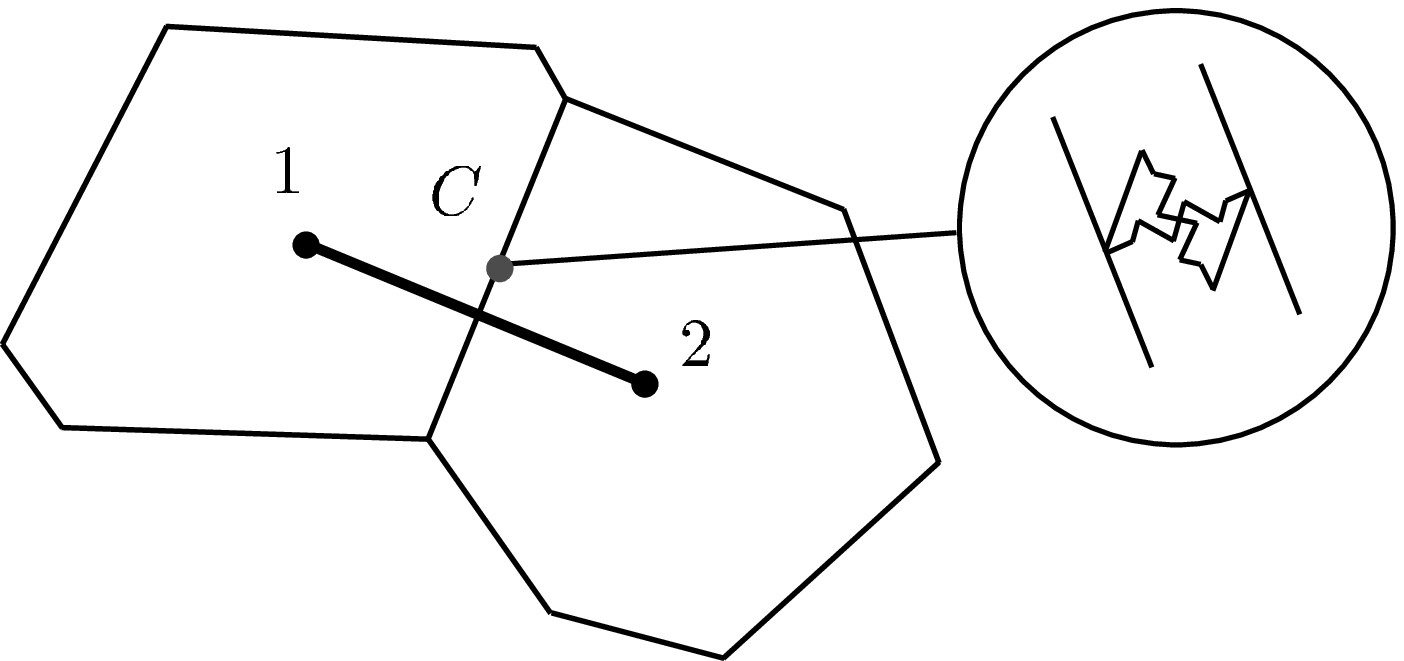, width=7cm} & \epsfig{file=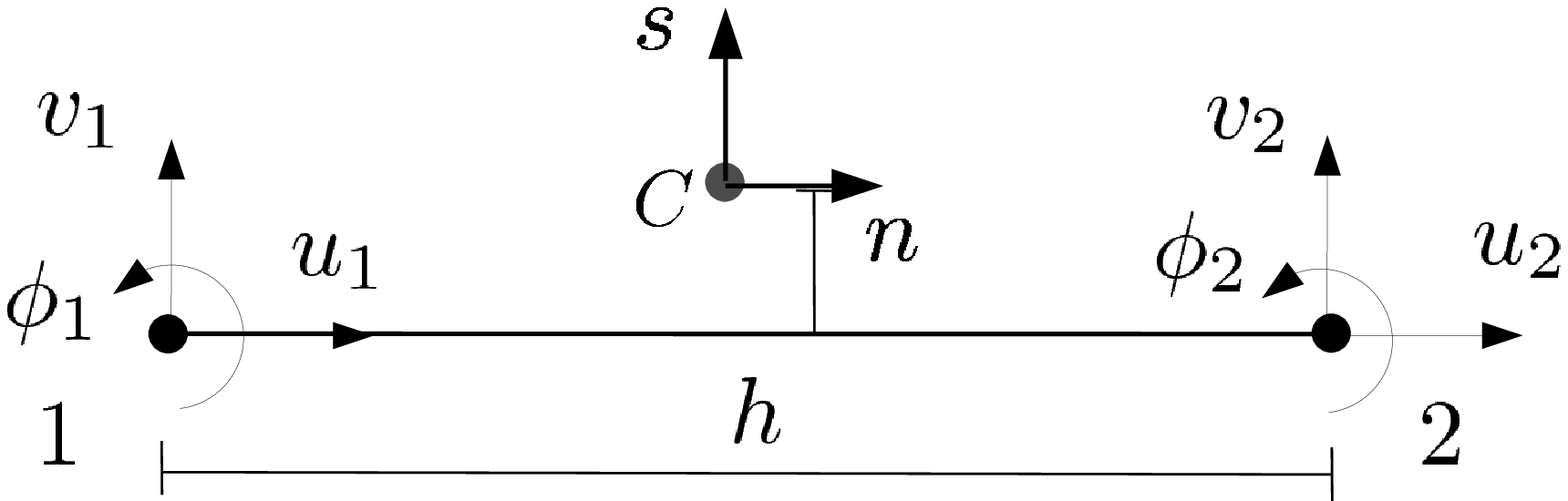, width=7cm}\\
(a) & (b)
\end{tabular}
\end{center}
\caption{a) Lattice element representing the connection between two nodes. b) Geometry in the local co-ordinate system.}
\label{fig:geomLat}
\end{figure}

The lattice elements describe the elastic and inelastic response of
concrete, i.e. both the elastic and inelastic displacements are
described by the displacement jump between rigid bodies. Each rigid
body possesses two translations and one rotation. In the local
coordinate system ($n$ and $s$), shown in Figure~\ref{fig:geomLat},
the degrees of freedom $\mathbf{u}_{\rm e} = \left\{u_1, v_1,
\phi_1, u_2, v_2, \phi_2\right\}^{\rm T}$ of two rigid bodies
sharing an interface describe the displacement discontinuities
$\mathbf{u}_{\rm c} = \left\{u_{\rm c}, v_{\rm c}\right\}^{\rm T}$
at the mid point $C$ of the interface, where the springs are located
by the relation
\begin{equation}
\mathbf{u}_{\rm c} = \mathbf{B} \mathbf{u}_{\rm e}
\end{equation}
where
\begin{equation} \label{eq:Bmatrix}
\mathbf{B} = \begin{bmatrix}
-1 & 0 & e & 1 & 0 & -e\\
0 & -1 & -h/2 & 0 & 1 & -h/2
\end{bmatrix}
\end{equation}
In Equation~(\ref{eq:Bmatrix}), $e$ is related to the eccentricity of the mid-point $C$ with respect to the element axis and $h$ is the length of the element (Figure~\ref{fig:geomLat}).
If $C$ is on the left hand side of the element, $e$ is positive.
Otherwise, $e$ is negative.
The displacements $\mathbf{u}_{\rm c}$ at the point $C$ are
transformed into strains $\sbf{\veps} = \left\{ \veps_{\rm n}
\hspace{0.2cm} \veps_{\rm s} \right\}^{T} = \mathbf{u}_{\rm c} /h$. 
The strain $\sbf{\veps}$ is related to
the stress $\sbf{\sigma} = \left\{\sigma_{\rm n} \hspace{0.2cm}
\sigma_{\rm s} \right\}^{T}$ by the constitutive model of the
springs, which is described in the following section.

The stiffness matrix of a lattice element in the local coordinate system results in
\begin{equation}
\mathbf{K}_{\rm t} = \dfrac{A}{h} \mathbf{B}^{\rm T} \mathbf{D}_{\rm t} \mathbf{B}
\end{equation}
where $A$ is the cross sectional area of the common interface of the two rigid bodies.

\subsection*{Constitutive model}\label{sec:const}
In the present study the response of concrete is analyzed with an interface model, which was developed in \cite{GraRem08} for cyclic loading and extended here to include thermal loading. In the following, the modeling approach is briefly reviewed to introduce the parameters, which are investigated in the present study. The interface model relies on stress based elasto-plasticity.
The strains are related to the stress $\boldsymbol{\sigma} = \left(\sigma_{\rm n}, \sigma_{\rm s}\right)^T$ as
\begin{equation} \label{eq:totStressStrain}
\boldsymbol{\sigma} = \mathbf{D}_{\rm e} \left(\boldsymbol{\varepsilon}_{\rm m} - \boldsymbol{\varepsilon}_{\rm p}\right)
\end{equation}
where $\mathbf{D}_{\rm e}$ is the elastic stiffness,  $\boldsymbol{\varepsilon}_{\rm p} = \left(\varepsilon_{\rm pn}, \varepsilon_{\rm ps}\right)^T$ is the plastic strain.
The mechanical strain $\sbf{\veps}_{\rm m}$ is defined as
\begin{equation}
\sbf{\veps}_{\rm m} = \sbf{\veps} - \sbf{\veps}_{\rm T}
\end{equation}
where the thermal strains are
\begin{equation} \label{eq:thermalStrain}
\sbf{\veps}_{\rm T} = \alpha_{\rm T} \left(T - T_{\rm a}\right) \begin{pmatrix} 1\\ 0 \end{pmatrix}
\end{equation}
The parameter $T$ is the temperature within the specimen and $T_{\rm a}$ is the reference temperature.
The thermal expansion coefficient $\alpha_{\rm T}$ is chosen as a
function of the temperature:
\begin{equation}
\alpha_{\rm T}\left(T\right) = \alpha_{\rm{T}1} + \alpha_{\rm{T}2} (T-T_{\rm a}) + \alpha_{\rm{T}3} (T-T_{\rm a})^2
\end{equation}
where $\alpha_{\rm{T}1}$, $\alpha_{\rm{T}2}$ and $\alpha_{\rm{T}3}$ are model parameters.

The elastic stiffness is
\begin{equation}
\mathbf{D}_{\rm e} = \begin{Bmatrix} E & 0\\
  0 & \gamma E
\end{Bmatrix}
\end{equation}
where $E$ and $\gamma$ are model parameters controlling both the Young's modulus and Poisson's ratio of the material \cite{GriMus01}.

\begin{figure} [h!]
  \begin{center}
    \epsfig{file=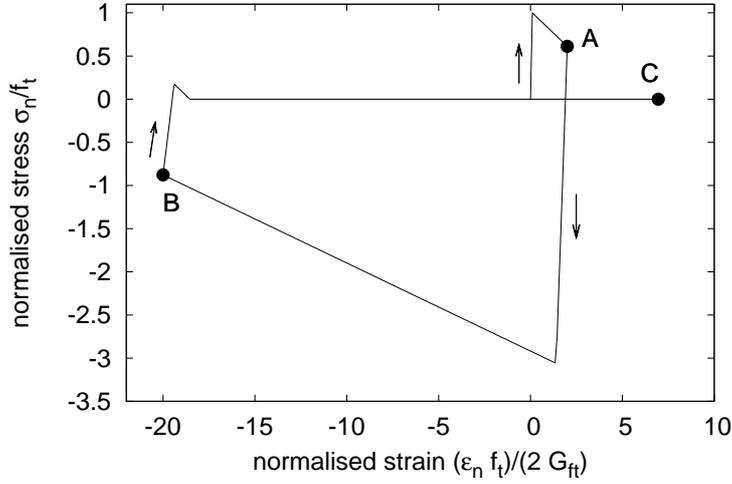, width=10cm}
  \end{center}
  \caption{Stress-strain response for fluctuating normal strains for  $\mu = 1$ (solid line) and $\mu = 0$ (dashed line).}
  \label{fig:constCyclic}
\end{figure}

The small strain plasticity model consists of the yield surface, flow rule, evolution law for the hardening parameter and loading and unloading conditions.
A detailed description of the components of the plasticity model is presented in \cite{GraRem08}.
The initial yield surface is determined by the tensile strength $f_{\rm t}$, by the ratio $s$ of the shear and tensile strength, and the ratio $c$ of the compressive and tensile strength.
The evolution of the yield surface during softening is controlled so that linear stress inelastic displacement laws for pure tension and compression are obtained, which are characterized by the fracture energies $G_{\rm ft}$ and $G_{\rm fc}$.
The eight model parameters $E$, $\gamma$, $f_{\rm t}$, $s$, $c$, $G_{\rm ft}$ and $G_{\rm fc}$ can be determined from a tensile, shear and compressive test of the material phase.
The constitutive response of the interface model is demonstrated by the stress-strain response for fluctuating normal strains in Figure~\ref{fig:constCyclic}.
The normal strain is increased to point $A$. Then the strain is reduced to point $B$ and again increased to point $C$.

\section*{Structural analysis}
The lattice modeling approach is applied to the meso-scale analysis
of the macroscopic phenomenon of transient thermal creep.
Experimental results performed by Thelandersson and reported in
\cite{The87} show that there is no significant influence of the
rate of temperature increase for rates of $1^\circ$C/min and
$5^\circ$C/min. This indicates that the temperature loading in these
experiments is slow enough to neglect the influence of the thermal
capacity of the materials. Therefore, the thermal analysis was
simplified by assuming a uniform temperature increase throughout the
specimen, disregarding the thermal conductivity and capacity, which
may differ significantly for the three phases and may play an
important role for increased heating rates.

\begin{figure} [h!]
\begin{center}
\begin{tabular}{cccc}
\epsfig{file=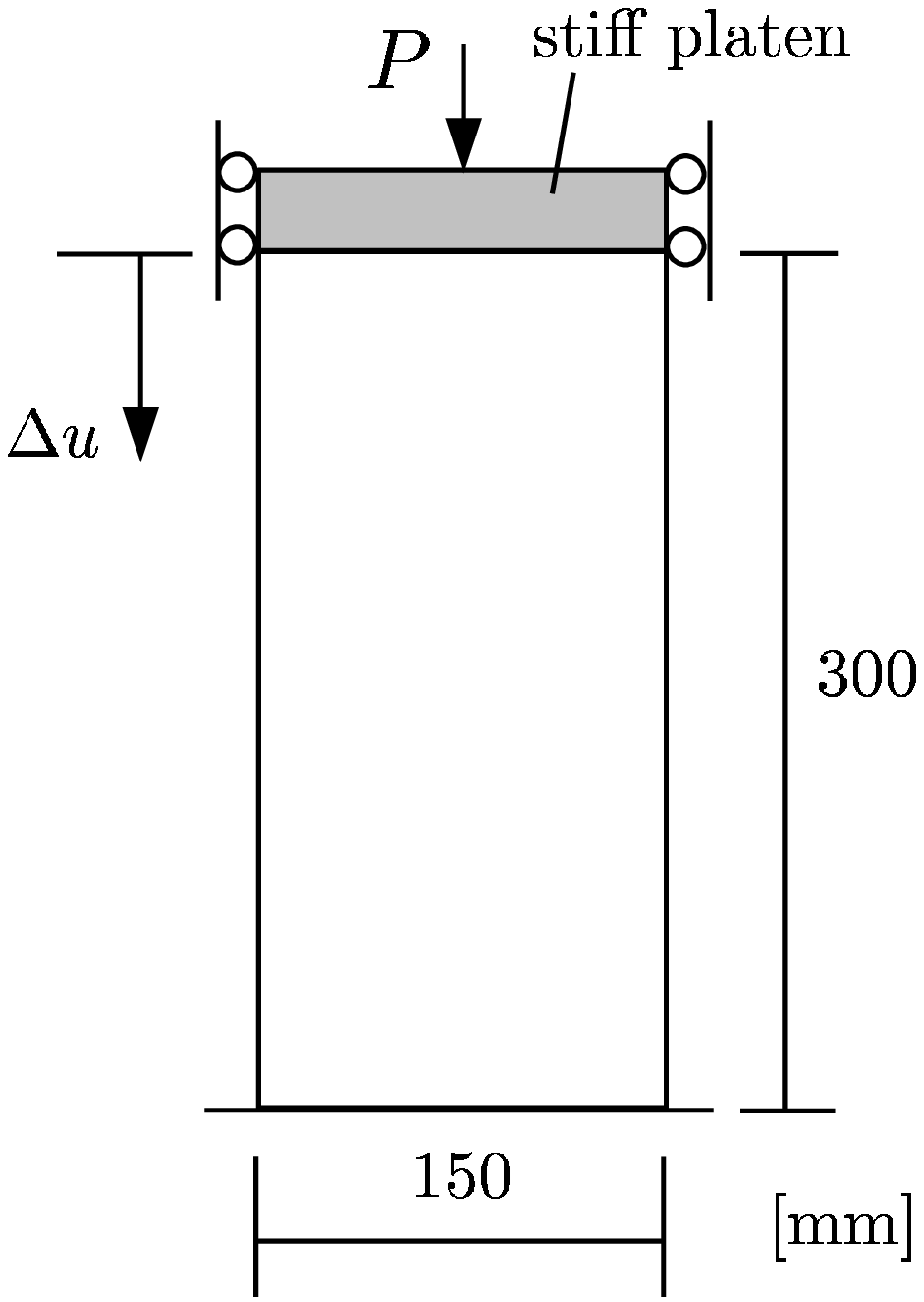, width= 5cm} & \epsfig{file=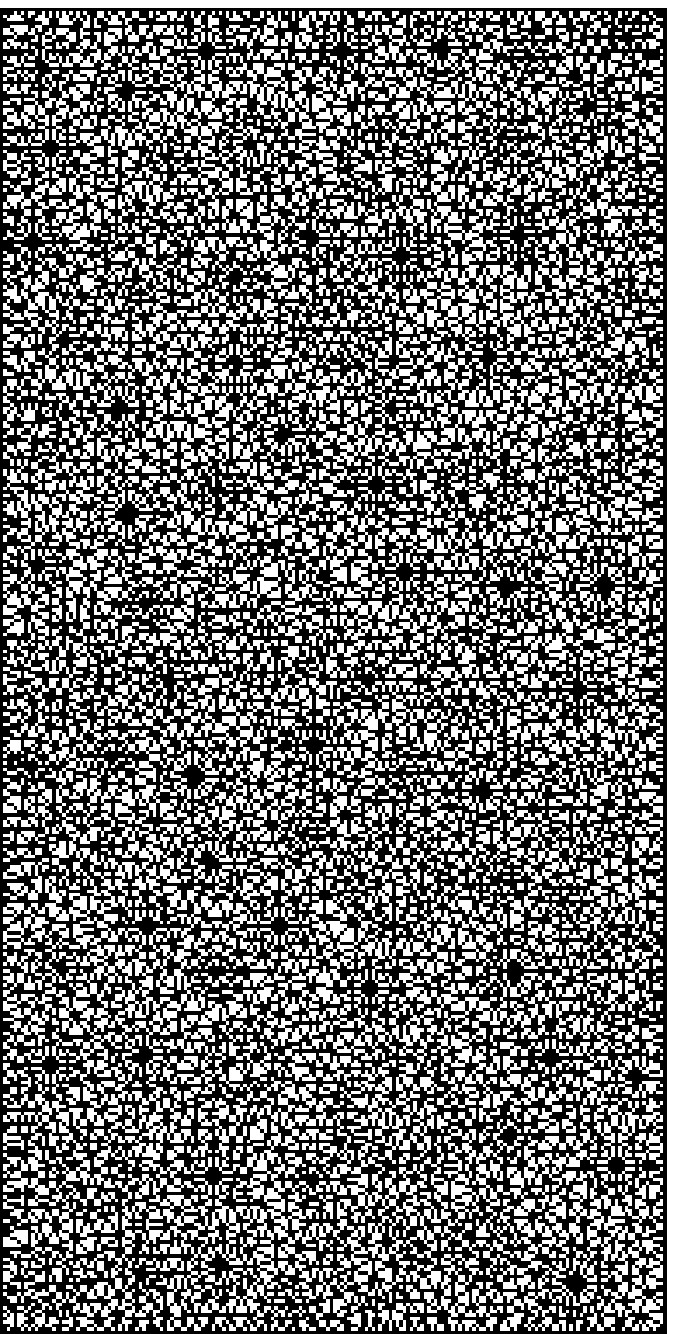, width= 3cm} & \epsfig{file=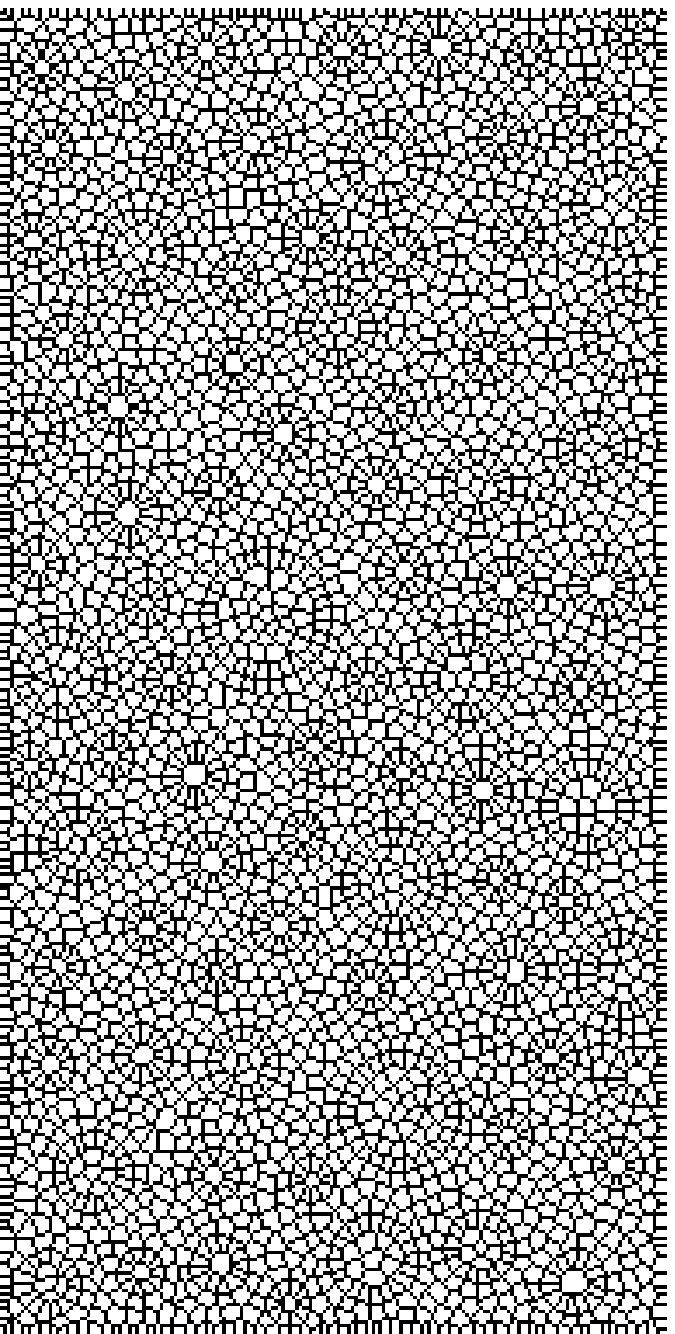, width= 3cm} & \epsfig{file=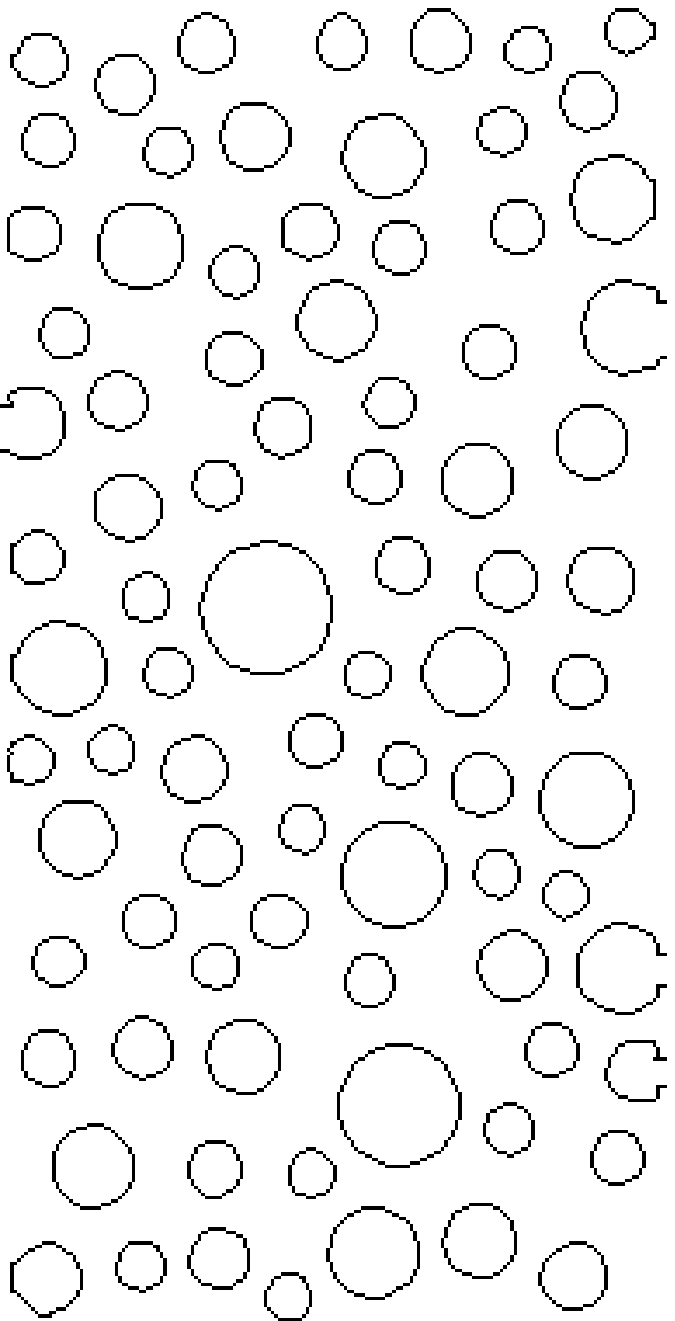, width= 3cm}\\
(a) & (b) & (c) & (d)
\end{tabular}
\end{center}
\caption{Model of a concrete prism: a) Geometry and loading setup,
b) Delaunay triangulation c) Voronoi tessellation d) Interfaces
representing the ITZ between matrix and aggregates.}
\label{fig:geom}
\end{figure}

A concrete prism was idealized by a two-dimensional discretization under the assumption of plane stress.
The height and width of the specimen was chosen as $L=300$~mm and $D=150$~mm, respectively.
The mesh was generated with a vertex density of $\rho = 1.5$ and a minimum distance of $d_{\rm m} = 3$~mm.
The distribution of aggregates distribution was obtained with $d_{\rm{max}} = 32$~mm, $d_{\rm{min}} = 10$~mm and an aggregate volume fraction of $\rho_{\rm a} = 0.3$. Aggregates of a diameter less than 10~mm are not modelled discretely.
The geometry, loading setup and mesh is shown in Figure~\ref{fig:geom}.

Three types of loading situations were investigated. Firstly, the concrete specimen was subjected to monotonic uniaxial compression at ambient temperature.
Secondly, the specimen was heated while subject to various levels of constant uniaxial compressive stress.
Thirdly, the specimen was subjected to uniform heating under uniaxial restraint.
The results of the latter two loading scenarios were compared to experimental results obtained by Thelandersson and reported in \cite{The87}.

\subsection*{Uniaxial compression}
The uniaxial compression analysis was controlled by the top displacement $\Delta u$. The average strain is determined as $\veps = \Delta u /L$.
The average stress $\sigma$ is determined as $P$/$D$, where $P$ is the reaction force caused by the top displacement $\Delta u$.

\begin{figure}[h!]
\begin{center}
\epsfig{file =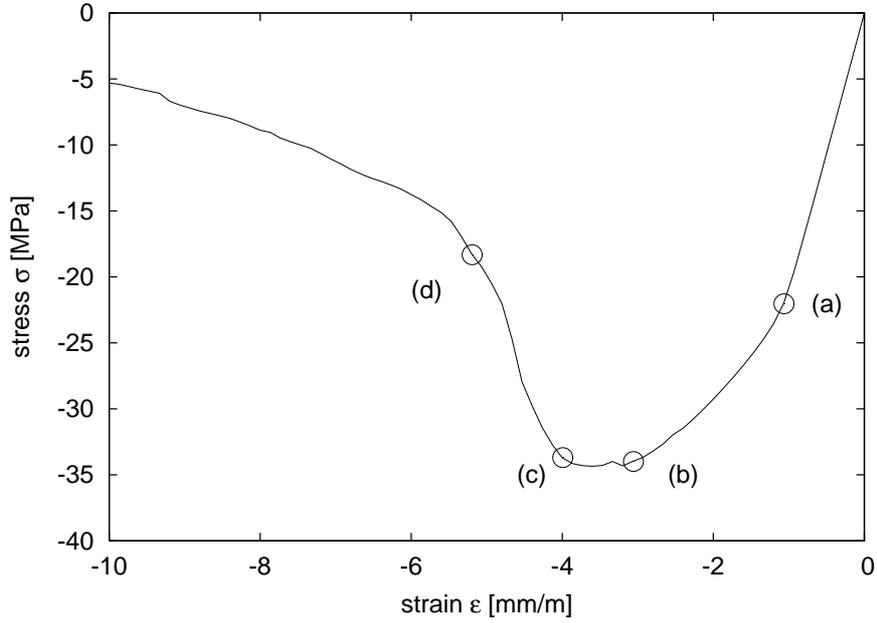,width=12cm}
\end{center}
\caption{Stress-strain response in uniaxial compression.}
\label{fig:strucEnv}
\end{figure}

\begin{figure}[h!]
\begin{center}
\begin{tabular}{cccc}
  \epsfig{file =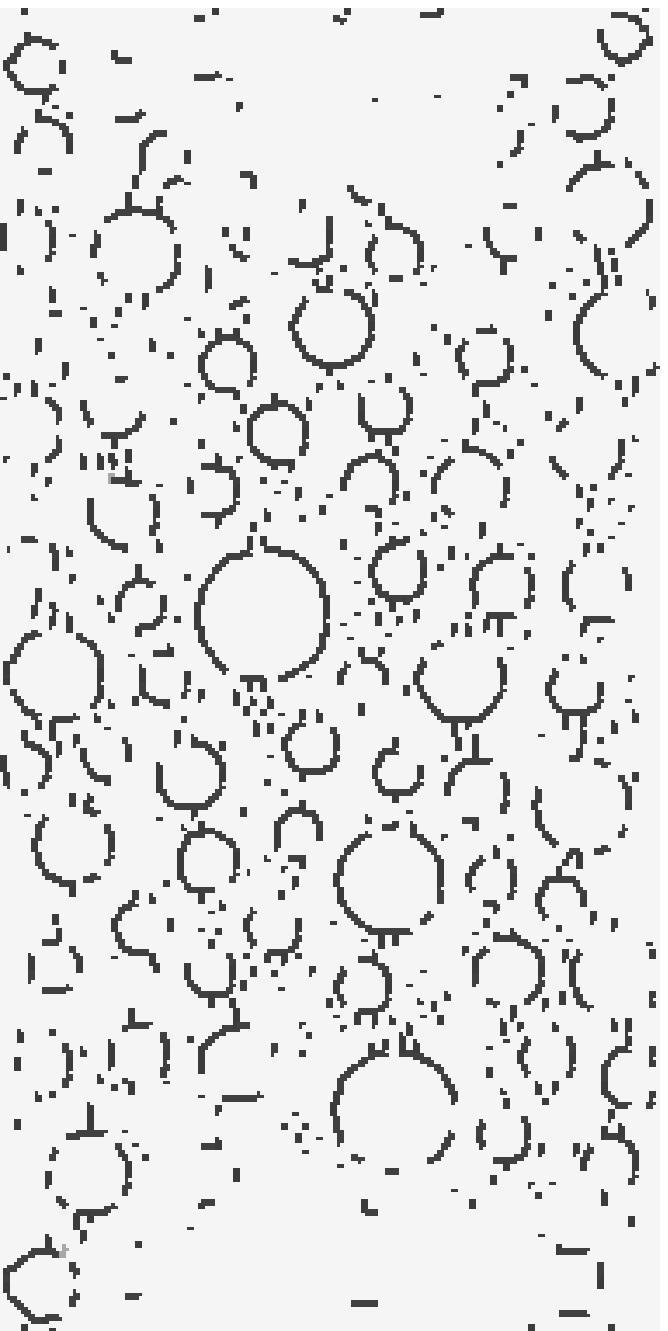, width=3cm} & \epsfig{file =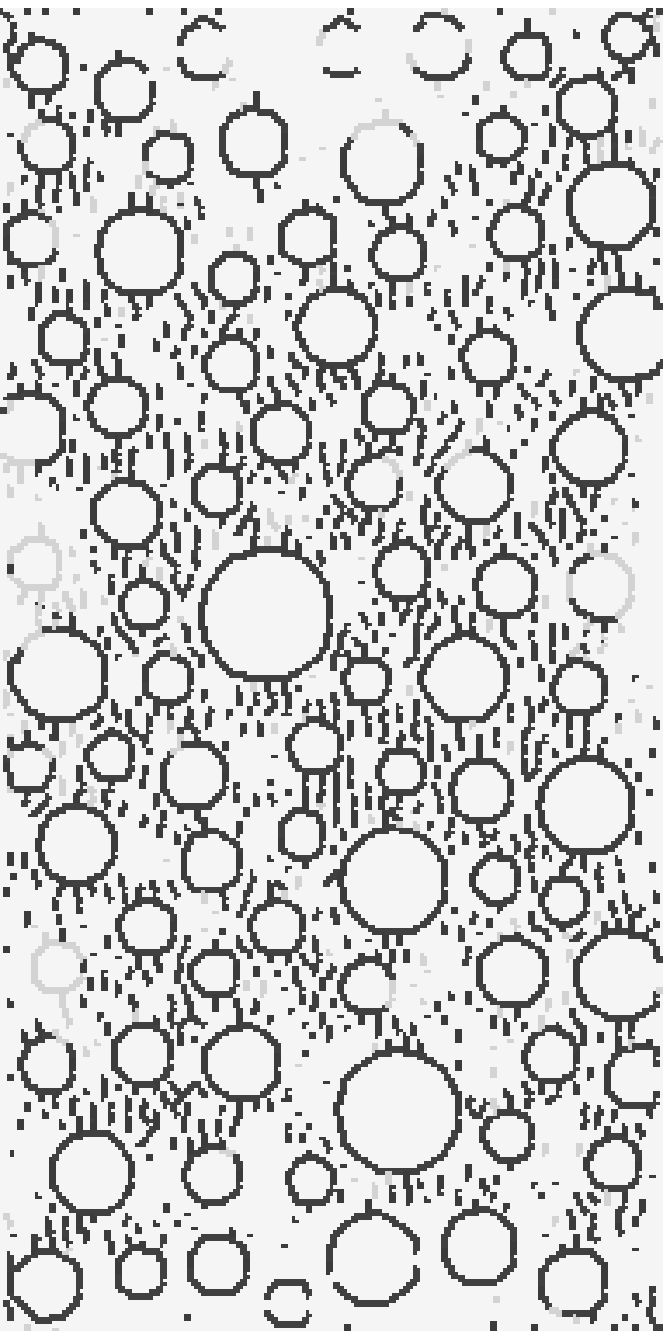, width=3cm} & \epsfig{file =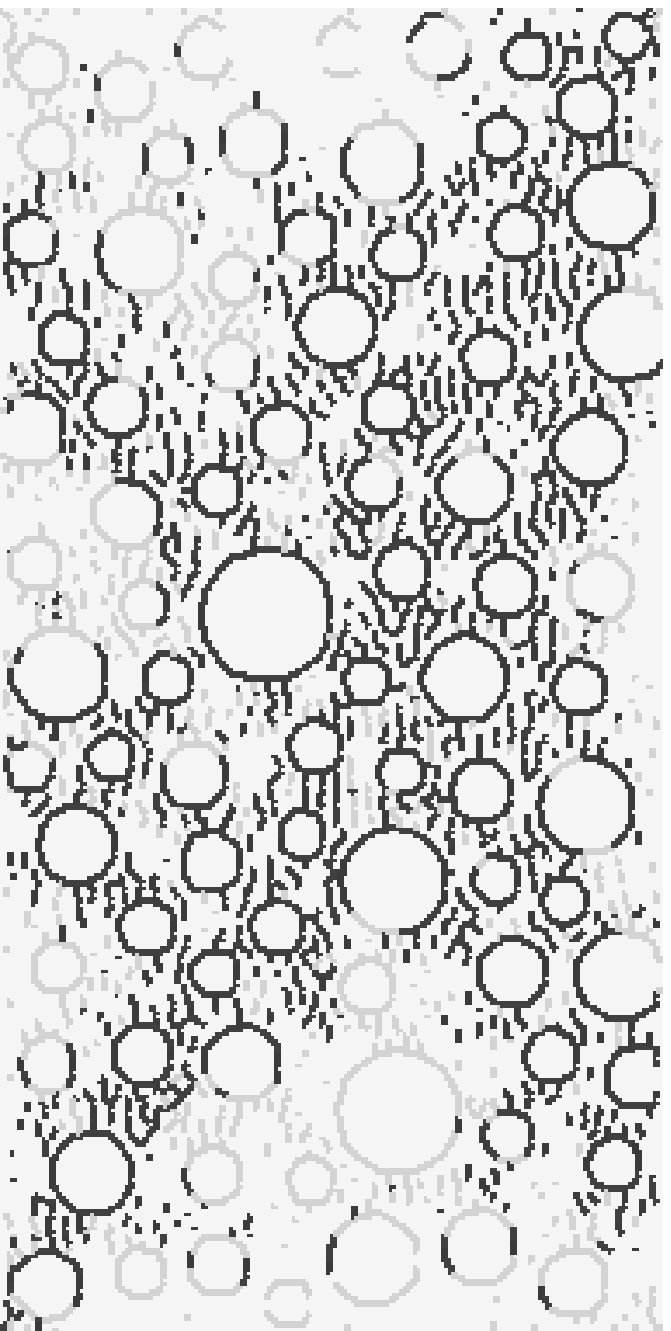, width=3cm} & \epsfig{file =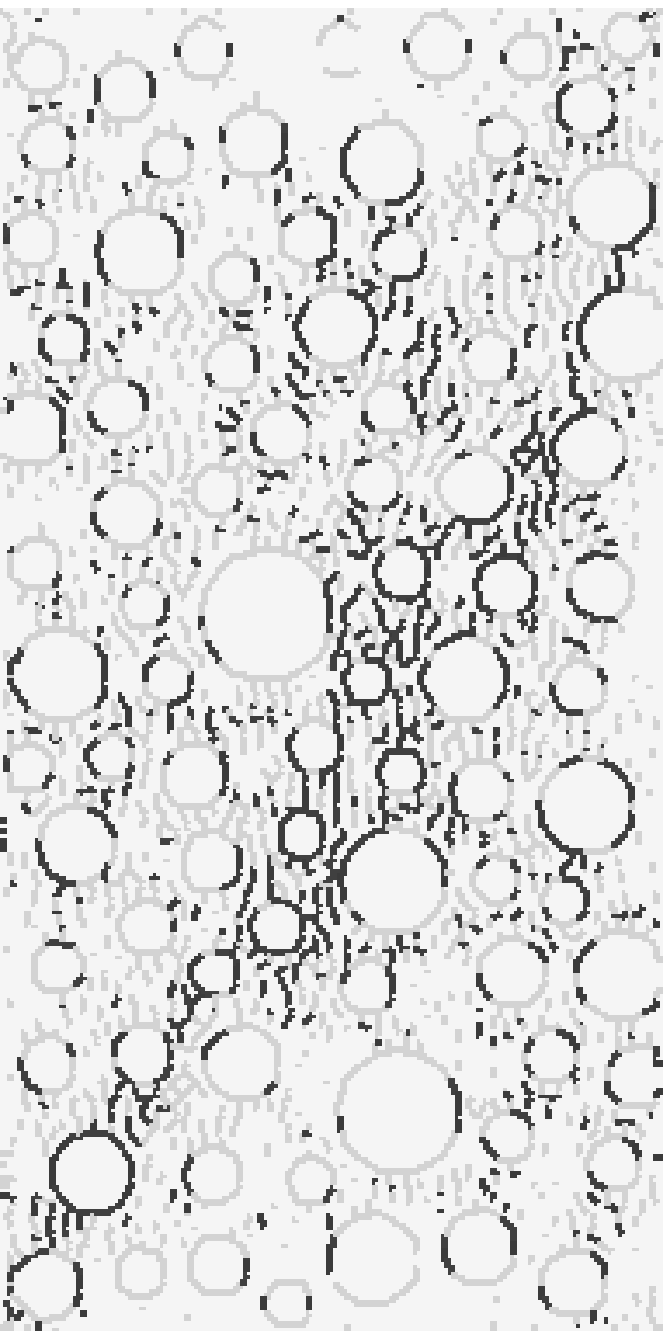, width=3cm}\\
(a) & (b) & (c) & (d)
\end{tabular}
\end{center}
\caption{The fracture patterns at four stages of analyses marked in \protect Figure~\ref{fig:strucEnv}.
  The black lines represent interfaces for which the damage variable increases.}
\label{fig:strucProcess}
\end{figure}

The model parameters for the mechanical part of the model are presented in Table~\ref{tab:strucParam}.
\begin{table}
\begin{center}
\caption{Model parameters for the structural analysis.}
\label{tab:strucParam}
\begin{tabular}{ccccccccc}
\hline\\
 & $E$ [GPa] & $\gamma$ & $f_{\rm t}$ [MPa] & $q$ & $c$ & $G_{\rm ft}$ [J/m$^2$] & $G_{\rm fc}$ [J/m$^2$] & $\mu$ \\
Matrix & $24$ & $0.33$ & $6$ & $2$ & $20$ & $600$ & $600000$ & $1$\\
ITZ & $34.3$ & $0.2$ & $1.5$ & $2$ & $20$ & $150$ & $150000$ & $1$\\
Aggregates & $60$ & $0.08$ & - & - & - & - & - & -\\
\hline\\
\end{tabular}
\end{center}
\end{table}
The parameters were chosen to obtain a macroscopic elastic stiffness of $E_{\rm m} = 21.5$~GPa, a compressive strength of $f_{\rm c} = 35$~MPa and an inelastic strain at peak of $\varepsilon_{\rm{i,max}}= 0.0015$, which were reported in \cite{The82}. The stress-strain response is shown in Figure~\ref{fig:strucEnv}.
The fracture patterns obtained at four stages of the analysis marked in Figure~\ref{fig:strucEnv} are shown in Figure~\ref{fig:strucProcess}.

At an early stage within the pre-peak regime, the cracks are mainly
restricted to the aggregate-matrix interface and are perpendicular
to the loading direction (Figure~\ref{fig:strucProcess}a). Around the peak load (Figures~\ref{fig:strucProcess}b~and~c), the interfaces between aggregates and matrix become strongly damaged and cracks occur in the matrix phase, which are orientated parallel to the direction of the compressive stress.
A similar direction of crack-propagation was observed in \cite{PicHelDor07} for the case of penny-shaped cracks embedded in an elastic matrix. 
At a later stage the cracks localize into a shear band
(Figure~\ref{fig:strucProcess}d), while the surrounding material
unloads.

The plasticity constitutive model limits the strength in
tension, shear and compression (Figure~\ref{fig:constCyclic}). In
the uniaxial compressive test, failure in the matrix is entirely
caused by tensile/low shear stress states. However, the compressive
strength in the ITZs is reached in the pre-peak regime of the
macroscopic load-displacement curve.

\subsection*{Heating under various compressive loads}
The second group of tests consists of two steps. First a uniaxial
compression stress is applied. Then, this stress is kept constant
and the temperature is increased from $20$~$^\circ$C to
$1000$~$^\circ$C, while the strain in the axial direction is
recorded. These tests are carried out for four stress levels of
$\sigma/f_{\rm c} = 0,0.225,0.45$ and $0.675$.

\begin{figure}[h!]
\begin{center}
\epsfig{file =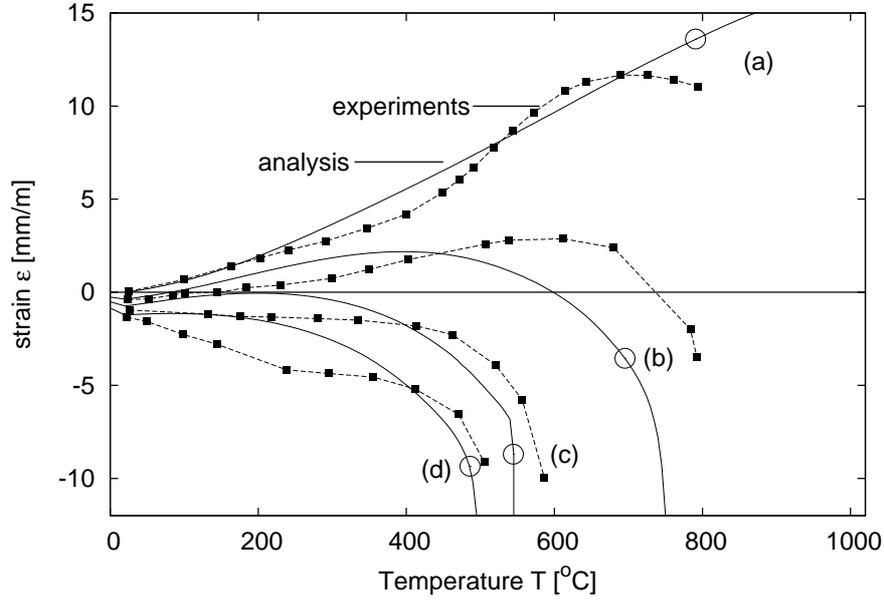,width=12cm}
\end{center}
\caption{Strain versus temperature for constant compressive stresses (a) $\sigma/f_{\rm c} = 0$, (b) $\sigma/f_{\rm c} = 0.225$, (c) $\sigma/f_{\rm c} = 0.45$ and (d) $\sigma/f_{\rm c} = 0.675$.}
\label{fig:strucStress}
\end{figure}

\begin{figure} [h!]
\begin{center}
\begin{tabular}{cccc}
 \epsfig{file =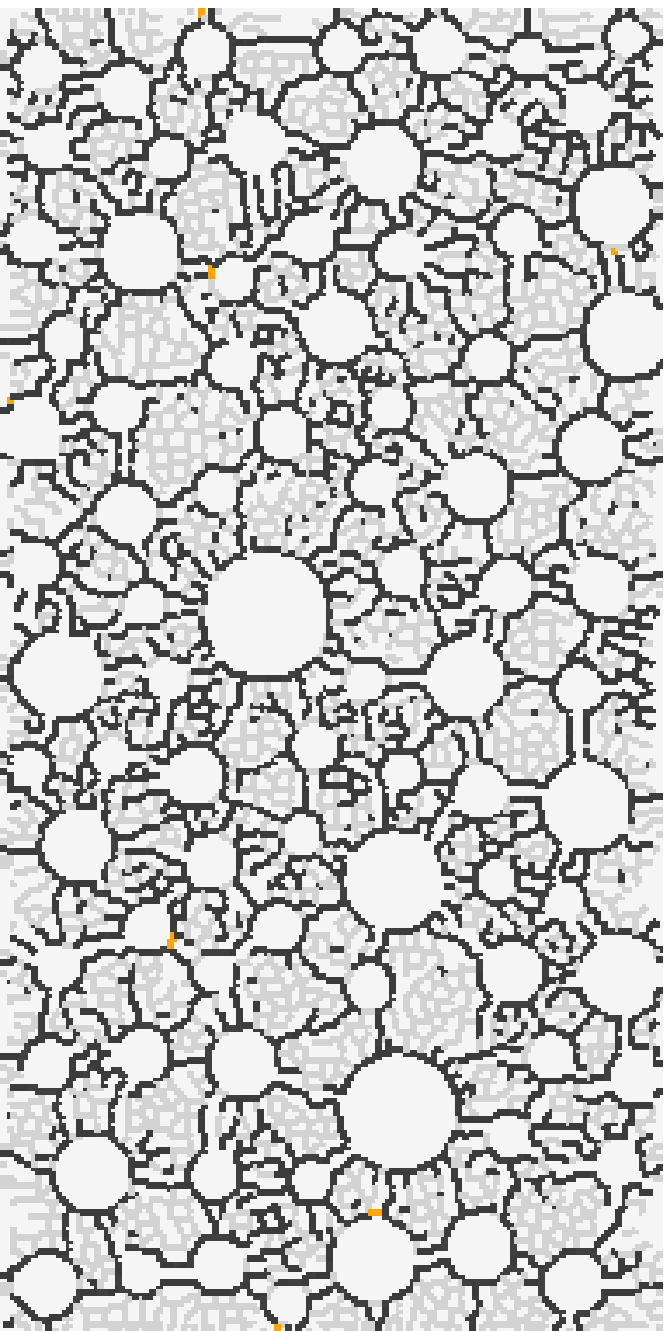, width=3cm} & \epsfig{file =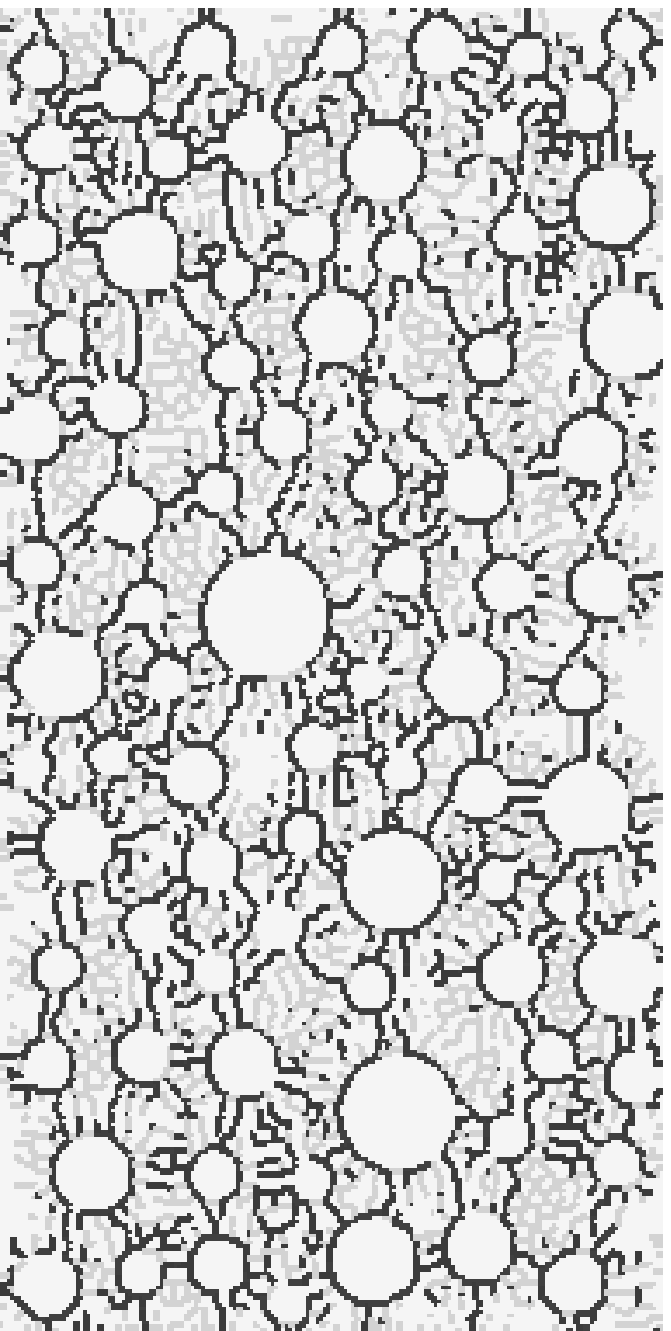, width=3cm} & \epsfig{file =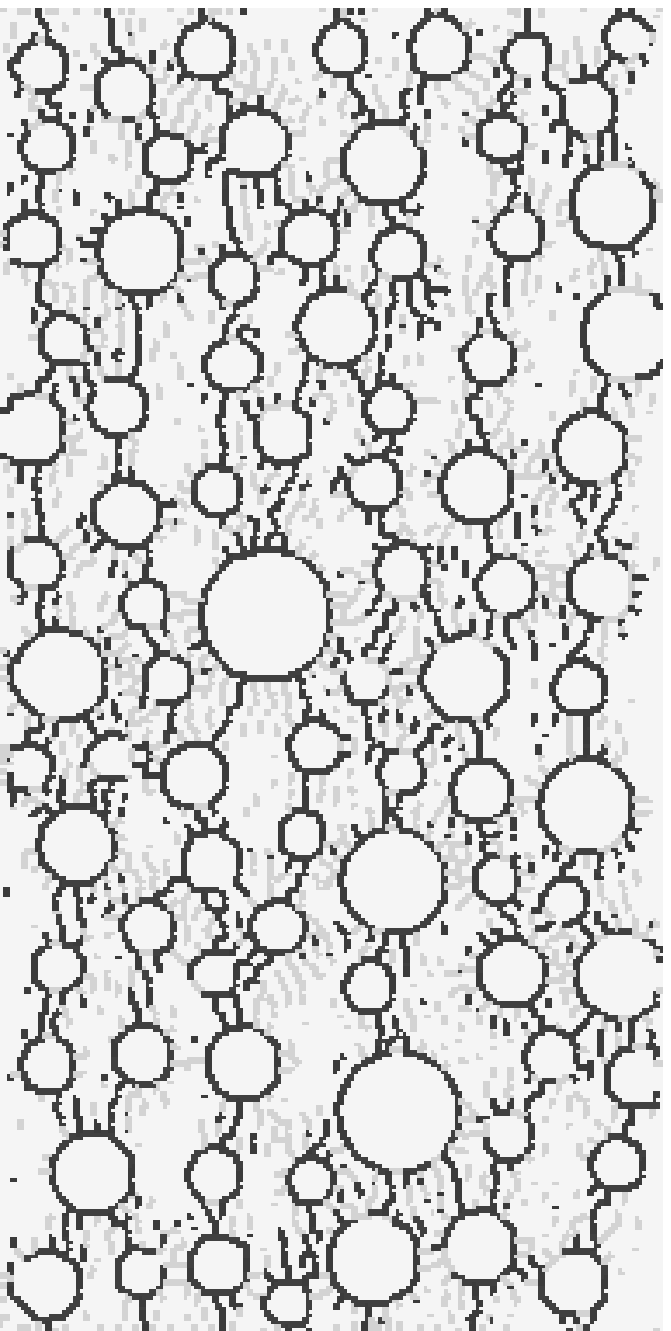, width=3cm} & \epsfig{file =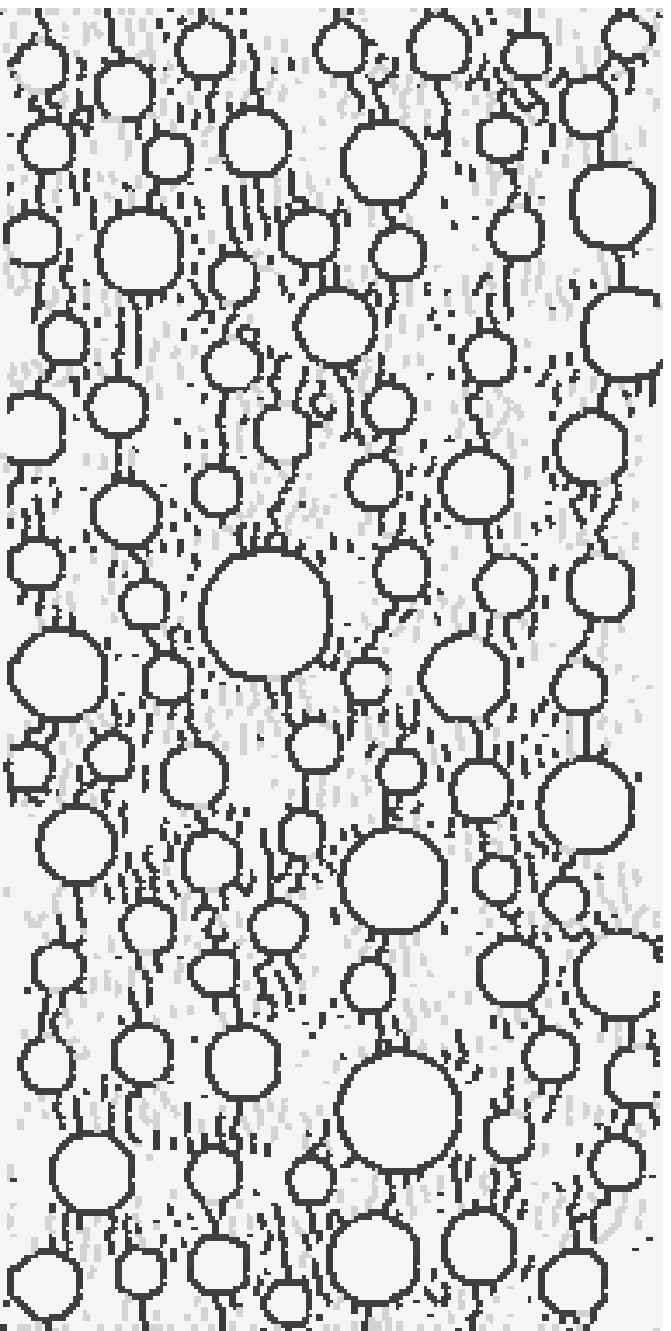, width=3cm}\\
(a) & (b) & (c) & (d)
\end{tabular}
\end{center}
\caption{The fracture patterns for (a) \protect $\sigma/f_{\rm c} = 0$, (b) \protect $0.225$, (c) \protect $0.45$ and (d) \protect $0.675$ at late stages of analyses marked in \protect Figure~\ref{fig:strucStress}.
  The black lines represent interfaces for which the damage variable increases.}
\label{fig:strucStressProcess}
\end{figure}

The thermal expansion coefficients for the matrix, interface and
aggregates are presented in Table~\ref{tab:strucThermal}. The Matrix
and ITZ phase initially expands and then contracts for higher
temperatures \cite{KhoGraSul85}, corresponding to dehydration of
the phases.
\begin{table}
\begin{center}
\caption{Thermal expansion coefficients for the structural analysis.}
\label{tab:strucThermal}
\begin{tabular}{cccc}
\hline
 & $\alpha_{\rm {T1}}$ [${}^\circ\textrm{C}^{-1}$] &  $\alpha_{\rm {T2}}$ [${}^\circ\textrm{C}^{-2}$] &  $\alpha_{\rm {T3}}$ [${}^\circ\textrm{C}^{-3}$]\\
Matrix & $1.5 \times 10^{-6}$ & $0$ & $-6\times 10^{-11}$\\
ITZ & $1.5 \times 10^{-6}$ & $0$ & $-6\times 10^{-11}$\\
Aggregates & $1.5\times 10^{-5}$ & $7 \times 10^{-8}$ & $0$\\
\hline\\
\end{tabular}
\end{center}
\end{table}
All other parameters are the same as in the previous section.
The thermal expansion coefficients of the three phases were determined from a fit of the model response to the experimental results for free expansion ($\sigma/f_{\rm c} = 0$). All other analyses are carried out with these parameters and no additional fits were performed.
Thus, the analysis for $\sigma/f_{\rm c} = 0.225$, $0.45$ and $0.675$ are model predictions.
The strain-temperature curves obtained with the meso-scale approach are compared to the experimental curves reported in \cite{The87} in Figure~\ref{fig:strucStress}.
The fracture patterns for the four different loading cases at a late stage of analysis marked in Figure~\ref{fig:strucStress} are shown in Figure~\ref{fig:strucStressProcess}.

For the case of free expansion ($\sigma/f_{\rm c} = 0$) cracks in
the matrix propagate radially outwards from the aggregate inclusions
due to the thermal incompatibility of the different phases. For the
other analyses, the crack orientations are influenced by the applied
compressive stresses. For a stress level of $\sigma/f_{\rm c} =
0.675$ the majority of the damaged interfaces in the matrix phase
are parallel to the loading direction.

\subsection*{Heating under uniaxial restraint}
\begin{figure} [h!]
\begin{center}
\epsfig{file =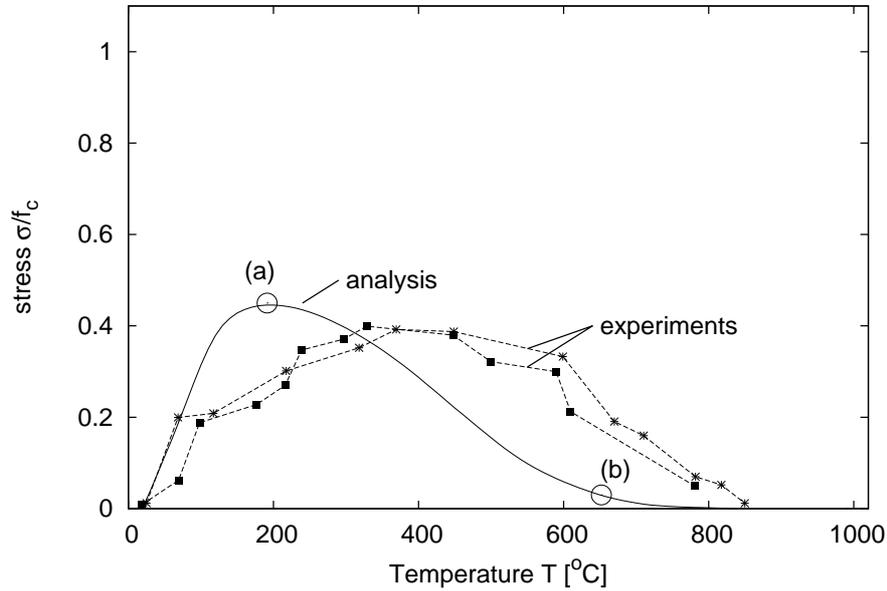,width=12cm}
\end{center}
\caption{Stress versus temperature for heating under uniaxial restraint.}
\label{fig:strucConf}
\end{figure}

\begin{figure} [h!]
\begin{center}
\begin{tabular}{cc}
  \epsfig{file =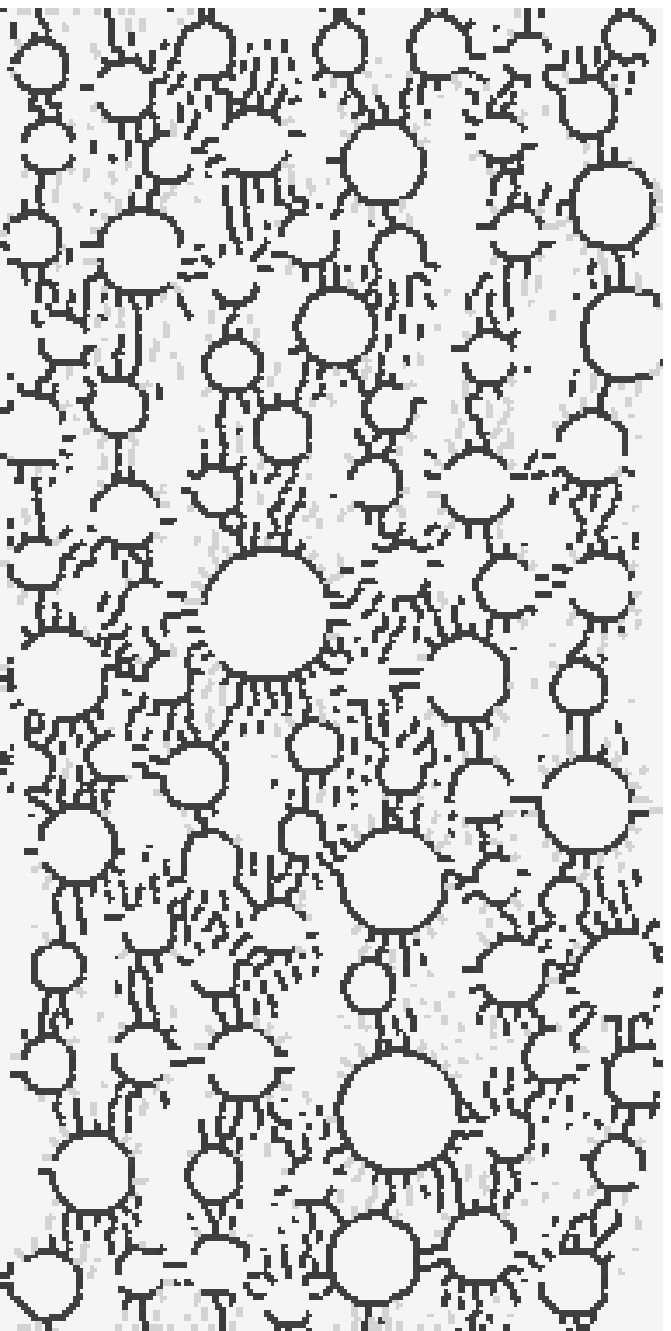, width=3cm} & \epsfig{file =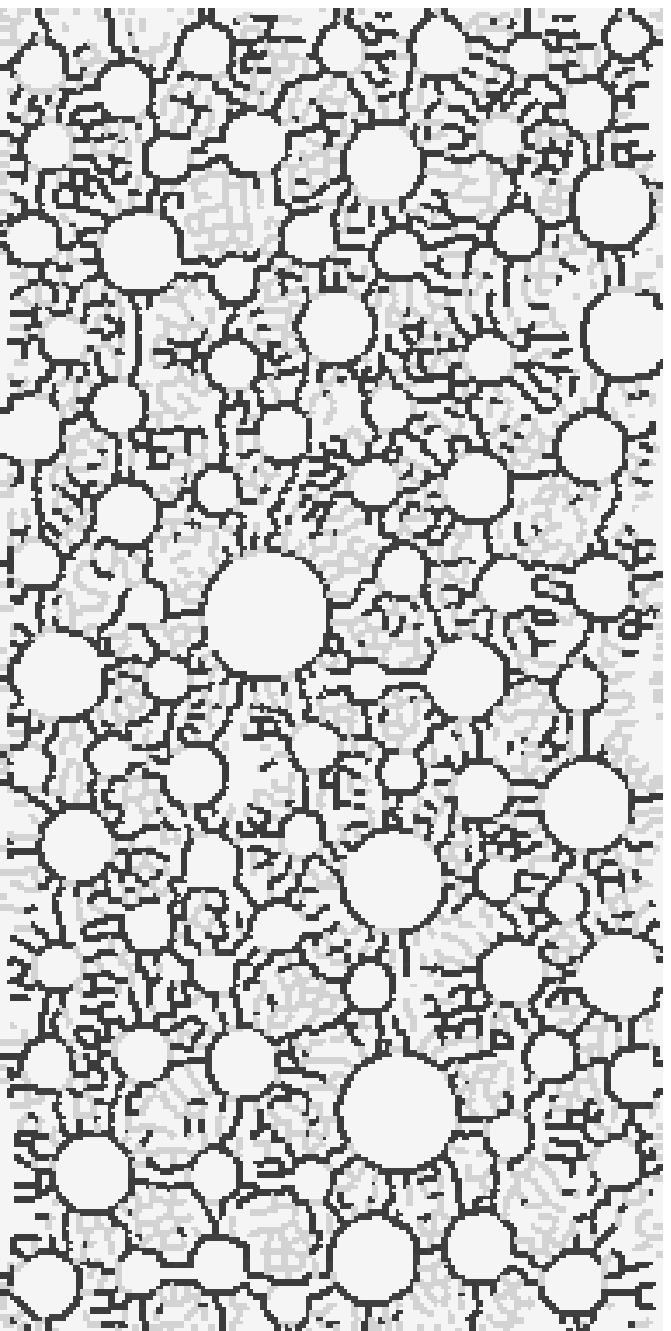, width=3cm}\\
(a) & (b)
\end{tabular}
\end{center}
\caption{The fracture patterns at two stages of analysis for heating under uniaxial restraint marked in \protect Figure~\ref{fig:strucConf}.
  The black lines represent interfaces for which the damage variable increases.}
\label{fig:strucConfProcess}
\end{figure}

The specimen is restrained in the axial direction and the
temperature is increased from the ambient temperature $20$~$^\circ$C
to $1000$~$^\circ$C. The same material properties as in the previous
example were used. The normalized average stress obtained from the
meso-scale analysis is plotted against temperature in
Figure~\ref{fig:strucConf} and compared with the experimental
results of Thelandersson reported in \cite{The87}. Initially, the
compressive stresses increase with increasing temperature until
approximately $200$~$^\circ$C and then reduce with further
temperature increase. It should be noted that the model parameters
adopted are not fitted to match the experimental results.
Nevertheless, the meso-scale modeling approach in this work
qualitatively captures the response observed in the experiments.
The fracture patterns at two stages of the analysis marked in Figure~\ref{fig:strucConf} are shown in Figure~\ref{fig:strucConfProcess}.

\subsection*{Macroscopic thermal damage}

\begin{figure} [h!]
\begin{center}
\epsfig{file =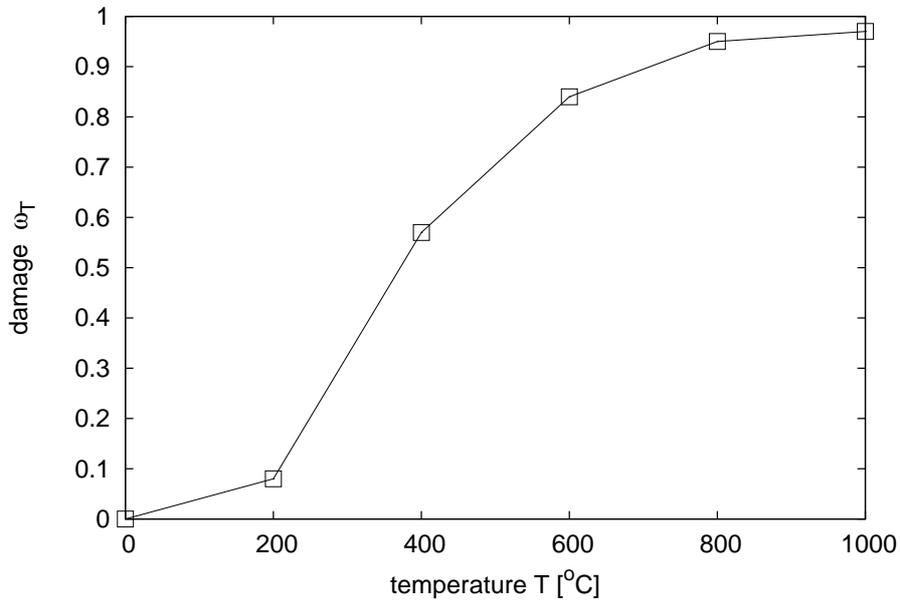,width=12cm}
\end{center}
\caption{Thermal damage versus temperature.}
\label{fig:temperatureDamage}
\end{figure}

Finally, the capabilities of the meso-scale approach are further illustrated by evaluating the stiffness at different stages of heating without any restraints.
The temperature is increased and a small uniaxial stress increment is applied  at different temperature levels to determine the stiffness of the material in the axial direction. The macroscopic stiffness $E_{\rm m}$ is computed as
\begin{equation}
E_{\rm m} = \dfrac{\Delta \sigma}{\Delta \veps}
\end{equation}
The stiffness obtained at ambient temperature is used as reference stiffness $E_0$ to compute the macroscopic thermal damage during heating as
\begin{equation}
\omega_{\rm T} = \dfrac{E_0 - E_{\rm m}}{E_0}
\end{equation}
The evolution of the macroscopic thermal damage during heating is presented in Figure~\ref{fig:temperatureDamage}.

\section*{Conclusions}

This paper presents a three-phase meso-scale approach to modelling concrete subjected to thermo-mechanical loading. An elasto-plasticity model for the mortar and interfacial transition zones is adopted that is able to describe the fracture process of concrete in compression. The model is also capable of qualitatively describing the main characteristics of concrete subjected to combined thermo-mechanical loading and the results indicate that the mismatch of thermal expansion has a strong influence on the thermal transient creep. Moreover, the path-dependence of the thermo-mechanical loading is described qualitatively correctly by the model. It is worth reiterating that, unlike other approaches, no phenomenological thermal damage mechanism is introduced to model the experimentally observed results, although phenomenological models for mechanical damage and thermal expansion of the individual phases are required. The present study is only a first step in a longer research project and further studies are required to investigate the influences of moisture and basic creep on thermal transient creep and the results will be presented in future publications.

\section*{Acknowledgments}
The simulations were performed with the object-oriented finite element package OOFEM \cite{Pat99,PatBit01} extended by the present authors. The mesh has been prepared with the mesh generator Triangle \cite{shew96}.

\end{document}